\documentstyle[12pt,epsf]{article}
\title{MAGNETIC CONFINEMENT AND SCREENING MASSES}
\author{ E.L.Gubankova$^{(a)}$ and  Yu.A.Simonov$^{(b)}$\\
(a)Institut f\"ur Theoretische Physik\\
Universit\"at Heidelberg\\
Philosophenweg 19,D-69120 Heidelberg,FRG\\
(b)Institute of Theoretical and
Experimental Physics\\
 117259, Moscow, B.Cheremushkinskaya 25,
Russia}
\date{}
\newcommand{\be}{\begin{equation}}
\newcommand{\ee}{\end{equation}}
\begin{document}
\maketitle

\begin{abstract}
Confined  and deconfined phases are defined through nonperturbative
correlators and nonperturbative background perturbation theory  is used
to compute the critical temperature $T_c$ and spatial string tension.
Taking evolution along one of the spatial axes the set of Hamiltonians
$H(n_1,n_2)$ with different Matsubara frequencies is obtained.

Meson and glueball screening masses and wave functions are computed
for $H(0,0)$ and compared with available lattice data.
\end{abstract}

1. Finite temperature QCD, and in particular the dynamics of
deconfinement phase
transition and the physics beyond the critical temperature $T_c$ is an
important testing  ground for our ideas about the nature of the QCD
vacuum.

During the last few years, a lot of lattice data have appeared which point
to the nonperturbative character of dynamics above $T_c$. To this
belong i) the area law of spatial Wilson loops [1,2], ii)the screening
 masses
of mesons and baryons [3-5] and glueballs [6], iii)the temperature
dependence of Polyakov--line correlators [7-9].
In addition,the behavior of $\varepsilon-3p$ above $T_c$ has a bump
incompatible with the simple quark--gluon  gas picture [10].
Thus the inclusion of nonperturbative $(NP) $ configurations into QCD
at $T>0$ and also at $T>T_c$ is necessary.

One of the  authors has developed a systematic method for QCD,
 treating $NP$
fields as a background and performing perturbative expansion around that
both for $T=0$ [11] and $T>0$ [12].

To describe the phase transition, a simple choice of deconfined phase
was suggested [12] where all  $NP$ color magnetic configurations are
kept intact as in the confined phase, whereas color electric
correlators that are responsible for confinement vanish.

This picture  together with  the background perturbation theory
[11,12] forms a basis of quantitative calculations, where field
correlators (condensates) are used as the $NP$ input.

As an immediate check one computes $T_c$ for
a different number of flavours $n_f$ in good agreement with lattice data.
One purpose of the present letter is to treat quantitatively
the points i)-iii) systematically in the framework of the same
method. Doing so we meet some limitations. First, in the description
of the deconfined phase in the spirit of the aforementioned method
[12], one uses the  magnetic correlators ( and spatial string
tension $\sigma_s$) from the lattice calculations [1,2]. In the
confined phase below $T_c$ and up to the temperatures $T\leq 1.5 T_c
$ in the deconfined phase  these correlators are  practically the same as
are known for $T=0$. Therefore in the mentioned temperature region the
spatial string tension $\sigma_s$ is regarded as $T$-- independent
 and coincides
with the standard string tension $\sigma$ at $T=0$
. This is because the main dynamics (and
contribution to the free energy) is due to the scale anomaly which
essentially changes at the temperature of the order of the dilaton
mass, i.e. at $T<1GeV$.  However, at large $T$, a new physics sets in:
that of dimensional reduction, which causes $\sigma_s$ to
grow [2,13].  For temperatures larger than 1.5 $T_c $  we have used
the scaling behavior of $\sigma_s(T)$ from [2], that is consistent
with the dimensional reduction regime. We  confine ourselves
to the region $0\leq T<2\div 3T_c$ and consider problems i)-iii)
point by point.

The second limitation is of technical character, since the proposed
 method is
used for simplicity in conjunction with the $1/N_c$ expansion. Therefore
values of $T_c$ are computed in the leading and subleading
 orders of $1/N_c$
expansion and strictly speaking the conclusion obtained for the first
 order of
deconfinement phase transition
cannot be  extended to $N_c=2$.

To calculate the screening masses one should first write the
 Green function of
quark--antiquark $(q\bar{q})$ or two gluons $(gg)$ in terms
 of the Wilson loop
average [14,15] and derive from it the Hamiltonian taking evolution
along one of the spatial axes. In fact, one obtains a set of Hamiltonians,
containing all higher Matsubara frequencies.  Our results for mesons
in general agree with one of the choices adopted before in [17], and
justify in that special case the confining ansatz made there.

Of special importance are the $gg$ Hamiltonian and the corresponding
 glueball
screening masses (GSM). Here the lowest Matsubara frequency is zero
 and the
next one is $2\pi T$.  Therefore within our picture only
the nonperturbative interaction between gluons -- spatial string tension
-- determine the temperature behavior of the lowest GSM. In contrast
to that, the meson screening masses for the lowest states have an
additional temperature dependence through the lowest
Matsubara frequency.

We calculate the GSM corresponding to lowest states and compare our
results with recent lattice
data [6]. With the scaling behavior $\sigma_s(T)$ [2] GSM forms the
similar pattern to that of the meson case.
 This picture is in agreement with existing lattice data
[7-9].

2. We derive here the basic formulas for the partition function,
 free energy and
Green  function in the nonperturbative background formalism at
$T>0$ [12]. The total gluonic field $A_{\mu}$ is split into a
perturbative part $a_{\mu}$ and a nonperturbative background $B_{\mu}$
\be
A_{\mu}=B_{\mu}+a_{\mu}
\ee
where both $B_{\mu}$ and  $a_{\mu}$ are subject to periodic boundary
conditions. The principle of this separation is immaterial for
 our purposes
here, and one can average over fields $B_{\mu}$ and $a_{\mu}$
 independently
using the 'tHooft's identity\footnote{private communication to one of
the authors (Yu.A.),December 1993.} (for  more discussion see [18])
\be
Z=\int DA_{\mu} exp (-S(A)) = \frac{\int DB_{\mu}\eta(B)\int Da_{\mu}
exp (-S(B+a))}{\int DB_{\mu}\eta(B)} \\ \ee $$
\equiv<<exp(-S(B+a)>_a>_B
$$
with arbitrary weight $\eta(B)$. In our case we choose $\eta(B)$ to fix
the field correlators and the string tension at their observed values. For
simplicity
have omitted gauge fixing and ghost terms in (2).

To the lowest order in $ga_{\mu}$ the partition function and free
energy are $$ Z_0=<exp(-F_0(B)/T)>_B, $$ \be F_0(B)/T=\frac{1}{2}ln
det W- lndet (-D^2(B))= \ee $$
=Sp\int^{\infty}_0\zeta(t)\frac{dt}{t}(-\frac{1}{2}e^{-tW}+e^{tD^2(B)})
$$
where $\hat{W}=-D^2(B)-2g\hat{F}$ and $D^2(B)$ is the inverse gluon and
ghost propagator, respectively, $\zeta(t)$ is a regularizing factor
[12].

The ghost propagator can be written as [11,12]
\be
(-D^2)^{-1}_{xy}=<x|\int^{\infty}_0 ds e^{sD^2(B)}|y>=
\int^{\infty}_0ds(Dz)^w_{xy}e^{-K}\Phi(x,y)
\ee
where standard notations [14,15] are used
$$K=\frac{1}{4}\int^s_0d\lambda\dot{z}^2_{\mu}~, ~~\Phi(x,y)=P exp
ig\int^x_yB_{\mu}dz_{\mu}$$
and a winding path integral is introduced [12]
\be
(Dz)^w_{xy}=\lim_{N\to
\infty}\prod^{N}_{m=1}\frac{d^4\zeta(m)}{(4\pi\varepsilon)^2}
\sum^{\infty}_{n=0,\pm1,..}
\int\frac{d^4p}{(2\pi)^4}e^{ip(\sum^{\infty}_{m=1}\
zeta(m)-(x-y)-n\beta\delta_{\mu 4})}
\ee
with $\beta=1/T$. For the gluon propagator an analogous expression holds
true, except that in (4) one should insert the gluon spin factor $P_F exp
2g\hat{F}$ into $\Phi(x,y)$. For a quark propagator the sum over windings
in (5) acquires a factor $(-1)^n$ and the quark spin factor is $exp
g\sigma_{\mu\nu}F_{\mu\nu}$ [12].

We are now in a position to make the expansion of $Z$ and $F$ in powers of
$ga_{\mu}$ (i.e. perturbative expansion in $\alpha_s$), and the
leading--nonperturbative term $Z_0, F_0$ -- can be represented as a sum of
contributions with different $N_c$ behavior of which we systematically will
keep the leading terms $0(N_c^2),0(N_c)$ and $0(N_c^0)$.

To describe the temperature of deconfinement phase transition one
 should specify
the phases and compute the associated free energy.
 For the confining phase to lowest
order in $\alpha_s$,the free energy is given by Eq.(3) plus contribution
of the energy density $\varepsilon $ at zero temperature
\be
F(1)=\varepsilon V_3-\frac{\pi^2}{30}V_3T^4-T\sum_s\frac{V_3(2m_s
T)^{3/2}}{8\pi^{3/2}}e^{-m_{s/T}}+0(1/N_c)
\ee
where $\varepsilon$ is defined by the scale anomaly [18]
\be
\varepsilon \simeq
-\frac{11}{3}N_c\frac{\alpha_s}{32\pi}<(F^a_{\mu\nu}(B))^2>
\ee
and the next terms in (6) correspond to the contribution of mesons (we
keep only pion gas) and glueballs. Note that $\varepsilon=0(N^2_c)$
while two other terms in (6) are $0(N^0_c)$.

For the high temperature or second phase we make an
assumption that there all color magnetic field correlators are the same as
in the first phase, while all color electric fields vanish. Since at
$T=0$ color magnetic correlators (CMC) and color electric
correlators (CEC) are equal due to the Euclidean $0(4)$ invariance,
one has
\be
<(F^a_{\mu\nu}(B))^2>=<(F^a_{\mu\nu})^2>_{el}+<(F^a_{\mu\nu})^2>_{magn};
<F^2>_{magn}=<F^2>_{el}
\ee

The string tension $\sigma$ which characterizes confinement is due to the
electric fields [20,21], e.g. in the plane (i4)
\be
\sigma=\sigma_E=\frac{g^2}{2}\int\int
d^2x<E_i(x)\Phi(x,0)E_i(0)\Phi(0,x)>+...
\ee
where dots imply higher order terms in $E_i$.

The vanishing of $\sigma_E$ implies that gluons and quarks are
liberated,and this will contribute
to the free energy in the deconfined phase via their closed loop terms
(3),(4) with all possible windings. The CMC enter via the perimeter
contribution $<\Phi(x,x)>\equiv \Omega$ (see (4)).  As a result one
has for the second high-temperature phase (cf.[12])  \be
F(2)=\frac{1}{2}\varepsilon
V_3-(N^2_c-1)V_3\frac{T^4\pi^2}{45}\Omega_g-\frac{7\pi^2}{180}N_cV_3T^4
n_f\Omega_q+0(N_c^0)
\ee
where $\Omega_q$ and $\Omega_g$ are perimeter terms for  quarks and
gluons, respectively,  the latter was estimated in [22] from the
adjoint Polyakov line.
 In what follows we replace
$\Omega$ by one for simplicity.

Comparing (6) and (10), and setting $F(1)=F(2)$ at $T=T_c$, one finds
to
order $0(N_c)$, disregarding all meson and glueball contributions
\be
T_c=\left(\frac{\frac{11}{3}N_c\frac{\alpha_s<F^2>}{32\pi}}{\frac{2\pi^2}{45}
(N^2_c-1)+\frac{7\pi^2}{90}N_cn_f}\right)^{1/4}
\ee
For the standard value of $G_2\equiv \frac{\alpha_s}{\pi}<F^2>=0.012
GeV^4$ [19] (note that for $n_f=0$ one should use an approximately 3
times larger value of $G_2$ [19]) one has for $SU(3)$ and different
values of $n_f=0,2,4$, respectively $T_c=~240,150,134.$ This should be
compared with lattice data [10] $T_c(lattice)=240,146,131$.The agreement
is quite good.  Note that at large $N_c$, one has $T_c=0(N_c^{0})$,
i.e. the resulting value of $T_c$ does not depend on $N_c$ in this
limit. Hadron contributions to $T_c$ are $0(N_c^{-2})$ and therefore
suppressed if $T_c $ is below the Hagedorn  temperature, as
typically happens [23].

3. In this section we derive the area law for spatial Wilson loops,
expressing spatial string tension in terms of CMC.

To this end we write $<W(C)>$ for any loop as [21]
\be
<W(C)>=exp[-\frac{g^2}{2}\int
d\sigma_{\mu\nu}(u)d\sigma_{\rho\lambda}(u')\ll
F_{\mu\nu}(u)\Phi(u,u')F_{\rho\lambda}(u')\Phi(u',u)\gg
\ee
$${\rm +\; higher\; order\; cumulants}]$$

For temporal Wilson loops, in the plane $i4, i=1,2,3,$ only color
electric fields $E_i=F_{i4}$ enter in (12), while  for spatial ones
in the plane $ik;i,k=1,2,3$ there appear color  magnetic fields
$B_i=\frac{1}{2} e_{ikl}F_{kl}$; in the standard way [21],
 one obtains the
area law for large Wilson loops of  size $L$, $L\gg T_g^{(m)}$
($T_g^{(m)}$ is the magnetic correlation  length)
 \be
<W(C)>_{spatial}\approx exp (-\sigma_s S_{\min})
\ee
where the  spatial string tension is [18]
\be
\sigma_s=\frac{g^2}{2} \int d^2x\ll B_n(x)\Phi(x,0)
B_n(0)\Phi(0,x)\gg+0(<B^4>)
\ee
 and $n$ is the component normal to the plane of the contour,
 while the last term in (14) denotes the contribution  of the fourth
 and higher order cumulants. On general grounds, one can write for the
 integrand in (14)
 \be
 \ll B_i(x)\Phi(x,0)B_j(0)\Phi(0,x)\gg=
 \delta_{ij}(D^B(x)+D_1^B(x)+\vec{x}^2\frac{\partial D^B_1}{\partial
 x^2})-x_ix_j\frac{\partial D^B_1}{\partial x^2},
 \ee
 and only the term $D^B(x)$ enters in (14) [18]
 \be
 \sigma_s=\frac{g^2}{2}\int d^2xD^B(x)+0(<B^4>)
 \ee
 This holds similarly for the temporal Wilson loop in the plane $i4$,so
that one  has
 the area law for $T<T_c$  with temporal string tension
 \be
 \sigma_E=\frac{g^2}{2} \int d^2xD^E(x)+0(<E^4>)
 \ee
 For $T=0$ due to the $0(4)$  invariance, CEC and CMC coincide and
 $\sigma_E=\sigma_s$.  For $T>T_c$ in the phase (2) CEC vanish, while
 CMC change on the scale of the dilaton mass $\sim 1 GeV$.Therefore
 one expects that $\sigma_s$ stays intact till the onset of the
 dimensional reduction mechanism. This expectation is confirmed by
 the lattice simulation -- $\sigma_s$ stays constant up to
 $T\approx 1.4 T_c$ [1,2].  Recent lattice data [2] show an increase
 of $\sigma_s$ at $T\approx 2T_c$ for SU(2),
 which could imply the early onset
 of dimensional reduction.

 4. In this section we consider the $q\bar{q}$ and $gg$ Green
 functions $G(x,y) $ at $T>T_c$ and derive corresponding Hamiltonians
 for evolution in the spatial direction. We start with the
 Feynman--Schwinger representation [24] for $G(x,y)$, where for
simplicity we
 neglect spin interaction terms \be
 G(x,y)=\int^{\infty}_0 ds \int^{\infty}_0 d\bar{s}
 e^{-K-\bar{K}}(Dz)^w_{xy}(D\bar{z})^w_{xy}<W(C)>
 \ee
 Here $K$ and $(Dz)^w_{xy}$ are defined in (5) and $<W(C)>$ in (12),
 where the contour $C$ is formed  by paths $z(\tau)$, $\bar{z}(\tau)$
 and $t\equiv x-y$ is for definiteness taken along the 3'rd axis.
 Since by definition at $T>T_c$ electric correlators are zero, only
 elements $d\sigma_{\mu\nu}$ in (12) in planes 12,13 and 23
 contribute. As a result one obtains for $<W(C)>$ the form (13) with
 \be
 S_{min}=\int^t_0
 d\tau\int^1_0
 d\gamma\sqrt{\dot{w}_i^2w_k^{'2}-(\dot{w}_iw'_i)^2}
 \ee
 where only spatial components $w_i, i=1,2,3$ enter
 \be
 w_i(\tau, \gamma)= z_i(\tau)\gamma +\bar{z}_i(\tau)(1-\gamma),
 \dot{w}_i = \frac{\partial w_i}{\partial \tau}~,~~
 w'_i=\frac{\partial w_i}{\partial\gamma}
 \ee
 The form (20) is equivalent to that used before in [14,15] but with
 $w_4\equiv 0$.

 As a next step one can introduce the "dynamical masses" $\mu,\bar{\mu}$
 similarly to [14,15]. We are looking for the "c.m" Hamiltonian which
 corresponds to the hyperplane where $z_3=\bar{z}_3$. Now the role of
 evolution parameter (time) is played by $z_3=\bar{z}_3=\tau$ with
 $0\leq \tau \leq t$, and we define transverse vectors
 $z_\bot(z_1,z_2),\bar{z}_\bot(\bar{z}_1,\bar{z}_2)$ and $z_4(\tau),
 \bar{z}_4(\tau)$.
 \be
  \frac{dz_3}{d\lambda}=\frac{d\tau}{d\lambda}=2\mu,~~
 \frac{d\bar{z}_3}{d\bar{\lambda}}=
\frac{d\tau}{d\bar{\lambda}}=2\bar{\mu}~~
 \ee
 Then $K,\bar{K}$ in (18) assumes the form
 \be
 K=\frac{1}{2}\int^t_0
 d\tau[\frac{m^2_1}{\mu(\tau)}+\mu(\tau)(1+\dot{z}^2_{\bot}+\dot{z}^2_4)]
 \ee
 and the  same for $\bar{K}$ with additional bars over
 $\mu,\dot{z}_{\bot}, \dot{z}_4$.

 Performing the transformation in the functional integral (18)
 $dsDz_3(\tau)\to D\mu,~~ d\bar{s}D\bar{z}_3(\tau) \to  D\bar{\mu}$,
 one has
  \be
   G(x,y)=\int
 D{\mu}D{\bar{\mu}}Dz_{\bot}
 D{\bar{z}_{\bot}}(D{z_4})^w_{xy}(D{\bar{z}_4})^w_{xy} exp (-A)
 \ee
 with the action
 \be
 A=K+\bar{K}+\sigma S_{min}
 \ee
 Note that $z_4,\bar{z}_4$ are not governed by NP dynamics and enter
 $A$ only kinematically   (through $K,\bar{K}$), and hence can be
 easily integrated out in (23) using Eq.(5) for  the 4-th
 components -- with $(x-y)_4=0$ (see [18] for more detail).

 One can now proceed as was done in [15,16], i.e. one introduces
 auxiliary functions $\nu(\tau,\gamma), \eta(\tau,\gamma)$,
 defines center-of-mass and relative coordinates $\vec{R}_{\bot},
 \vec{r}_{\bot}\equiv \vec{r}$,
  and finally integrates out $\vec{R}_{\bot}$ and
  $\eta(\tau,\gamma)$. The  only difference from [15,16] is that now
  $z_4,\bar{z}_4$ do not participate in all those transformations. As
  a result one obtains
  \be
  G(x,y)\sim \int D\nu D\mu D\bar{\mu}
  Dre^{-A^{(1)}}\sum_{n,n_2}e^{-A_{n_1n_2}}
\ee
here
  \be
  A^{(1)}(\mu,\bar{\mu},\nu) = \frac{1}{2} \int^t_0
  d\tau[\frac{\vec{p}^2_r+m^2_1}{\mu}
+\frac{\vec{p}^2_r+m^2_2}{\bar{\mu}}+\mu+\bar{\mu}
  +\sigma^2r^2\int^1_0\frac{d\gamma}{\nu}+\int^1_0\nu d\gamma+
  \ee
  $$
  +\frac{\vec{L}^2/r^2}{\mu(1-\zeta)^2+\bar{\mu}\zeta^2+\int^1_0
  d\gamma(\gamma-\zeta)^2\nu}]
  $$
  \be
  A_{n_1n_2}=\frac{1}{2}(\pi T)^2\int_0^t
  d\tau(\frac{b^2(n_1)}{\mu(\tau)}+\frac{b^2(n_2)}{\bar{\mu}(\tau)}),
  \ee
  $b(n)=2n$ for bosons and $2n+1$ for quarks. We also have introduced
  the radial momentum $\vec{p}_r$,the angular momentum $\vec{L}$
  \be
  \vec{p}^2_r\equiv\frac{(\vec{p}\vec{r})^2}{r^2}=
  (\frac{\mu\bar{\mu}}{\mu+\bar{\mu}})^2
  \frac{(\vec{r}\dot{\vec{r}})^2}{r^2},~~\vec{L}=\vec{r}\times
  \vec{p}
  \ee
  and
  \be
  \zeta(\tau)=\frac{\mu(\tau)+<\gamma>\int\nu
  d\gamma}{\mu+\bar{\mu}+\int\nu d\gamma}, ~~<\gamma>\equiv
  \frac{\int\gamma\nu d\gamma}{\int\nu d\gamma}
  \ee

Let us  define
the Hamiltonian $H$ for the given action $A=A^{(1)}+A_{n_1n_2}$ in (25),
integrating over $D{\nu},D{\mu},D{\bar{\mu}}$ around the extremum of
$A$ ( this is an exact procedure in the limit $t\to \infty$).
For the extremal values of the auxiliary fields, one has
$$
\frac{\vec{p}^2+m^2_1+(b(n_1)\pi T)^2}{\mu^2(\tau)} = 1
-\frac{l(l+1)}{\vec{r}^2}(\frac{(1-\zeta)^2}{a^2_3}-\frac{1}{\mu^2})$$
 \be
\frac{\vec{p}^2+m^2_2+(b(n_2)\pi T)^2}{\bar{\mu}^2(\tau)} = 1
-\frac{l(l+1)}{\vec{r}^2}(\frac{\zeta^2}{a^2_3}-\frac{1}{\bar{\mu}^2})
\ee
$$
\frac{\sigma^2}{\nu^2(\tau,\gamma)}\vec{r}^2=
1-\frac{l(l+1)}{\vec{r}^2}\frac{(\gamma-\zeta)^2}{a^2_3}$$
where
$a_3=\mu(1-\zeta)^2+\bar{\mu}\zeta^2+\int
d\gamma(\gamma-\zeta)^2\nu$ and $\zeta$ is
defined  by eq.(29).

After the substitution of these extremal
values into the path integral Hamiltonian
   \be
    G(x,y) = <x|\sum_{n_1n_2}e^{-H_{n_1n_2}t}|y>
    \ee
one has to construct (performing proper Weil
ordering  [28]) the operator Hamiltonian
acting on the wave functions.

Consider for simplicity the case $\vec{L}=0$.Then from (30), one
obtains

 \be
  H_{n_1n_2}= \sqrt{\vec{p}^2+m^2_1+(b(n_1)\pi T)^2}+
\sqrt{\vec{p}^2+m^2_2+(b(n_2)\pi T)^2}+ \sigma r
\ee
Here $\vec{p}=
\frac{1}{i}\frac{\partial}{\partial\vec{r}}$ and
$\vec{r}=\vec{r}_{\bot}$ is a $2d$ vector,
$\vec{r}=(r_1,r_2);~m_1,m_2$ -- current masses of
quark and antiquark and $\sigma=\sigma_s$, for the $gg$ system $m_1=m_2=0$
and $\sigma=\sigma_{adj}=\frac{9}{4}\sigma_s$for SU(3) [25].

Eigenvalues and eigenfunctions of $H_{n_1n_2}$
\be
H_{n_1n_2}\varphi(r)=M(n_1,n_2)\varphi(r)
\ee
define the so--called screening masses and corresponding
wave--functions, which have been measured in lattice calculations
[3-6].

The lowest mass sector for mesons is given by $H_{00}$, where for
$n_1=n_2=0$ one has $b(n_1)=b(n_2)=1$ in (32). For light quarks one
can put $m_1=m_2=0$ and expand at large $T$ square roots in (32) to
obtain
\be
 H_{00}\approx 2\pi T+ \frac{\vec{p}^2}{\pi T}+ \sigma_s r
,~~ M_{00}\approx 2\pi T + \varepsilon (T)
\ee
 where $\varepsilon(T)=\frac{\sigma_s(T)^{2/3}}{(\pi T)^{1/3}}a $
 and $a\simeq 1.74$ is the eigenvalue of the 2-D dimensionless
 Schr\"{o}dinger equation.

 Assuming the parametrization $\sqrt{\sigma_s} =cg^2(T)T$ with
c being numerical constant different for $SU(2)$ and $SU(3)$ [2]
 and the scaling behavior of $g^2(T)$,
 one has $M_{00}\approx 2\pi T+O((\ln(T/\Lambda_T))^{-4/3}T)$ tending
to twice the lowest Matsubara frequency,as was suggested before in
[26].(This limit corresponds to the free quarks,propagating
perturbativly in the space--time with the imposed antiperiodic boundary
conditions along the 4th axis).
Eq. (34) coincides with that proposed in [17],where also numerical study
was done of $M_{00}$ and $\varphi_{00}(r)$.Our calculations of Eq.(33-34)
for $SU(3)$ with $c=0.586$ [2] agree with [17]
 and are presented in Fig.1 together with lattice
calculations of $\varphi(r)_{00}$ for $\rho$--meson [5].
The values of $M_{00}(T)$ found on
the lattice [3,4] are compared with our results in Fig.2.  Note, that
our $M_{00}$ (34) does not contain perimeter corrections which are
significant.  Therefore one has to add the meson
 constant to $M_{00}$ to compare with lattice data.  We
disregard the spin--dependent and one-gluon-exchange (OGE) interaction
here for lack of space and will treat this subject elsewhere [18].
It is known [17], however, that OGE is not very important at around
$T\approx 2T_C$.

We note that the lowest meson screening mass  appears also in Yukawa
type exchanges between quark lines and can be  compared with the
corresponding masses in Yukawa potentials.

 5.  For the $gg$ system, the lowest
mass sector is given by $b(n=0)=0$, and one has from (32) \be
H_{00}(gg)=2|\vec{p}|+\sigma_{adj}r \ee
 Note that $T$ does not enter
the kinetic terms of (35).

To calculate with (35) one can use  the approximation in (26) of
$\tau$--independent $\mu$ [14], which leads to the operator
$(\mu=\bar{\mu})$
 \be h(\mu)=\mu+\frac{\vec{p}^2}{\mu}+\sigma_{adj}r
\ee
The eigenvalue $E(\mu)$ of $h(\mu)$ should be minimized with respect to $\mu$
and the result $E(\mu_0)$ is known to yield the eigenvalue of (35)
within a few percent accuracy [27]. The values $E(\mu_0)\approx
M_{gg}$ thus found are  presented  in Fig.2. The
corresponding wave functions $\varphi_{00}(r)$ are given in Fig.1.
These data can be compared with the glueball screening masses, found on the
lattice in [6].

Another point of comparison is the Polyakov line correlator
\be
P(R)=\frac{<\Omega^+(R)\Omega(0)>}{<\Omega^+(R)><\Omega(0)>}= exp(-V(R)/T)
\ee
 with $$V(R)=\frac{exp(-M
R)}{R^{\alpha}}~,~~\Omega(R)=\frac{1}{N_C} tr P exp( ig\int^{\beta}_0
A_4dz_4) $$
 It is easy to understand that $M=M_{gg}$ and one can
compare our results for the GSM $M_{gg}(T)$ with the corresponding
lattice data for  $M$ in Fig.2.

 Taking  again into account an unknown constant perimeter correction to our
 values of $M_{gg}$, one can see a reasonable qualitative behavior.
 We refer the reader to another publication for more detail [18].

In addition to the purely nonperturbative source of GSM described above
there is a competing mechanism
 -- the perturbative formation of the electric Debye
screening mass $m_{el}(T)=gT(\frac{N_c}{3}+\frac{n_f}{6})$ for each
gluon with $g(T)$ given by the temperature dependence of
 $\sqrt{\sigma_s(T)}=c g^2(T)T$
[2]. Therefore for large $T$, where $m_{el}$ is essential, one should
rather use instead of (35) another Hamiltonian, which is obtained
from (32) replacing $m_1=m_2=m_{el}(T)$ \be
\tilde{H}_{00}(gg)=2\sqrt{\vec{p}^2+m^2_{el}(T)}+\sigma_{adj}(T)r
\ee
It turns out that up to temperatures  $T\leq 2 T_c$, one can
consider the effect of $m_{el}(T)$  as a small
correction to the eigenvalue of (35) $E$
\be
M_{gg}=  E+\delta E\approx 4(\frac{a}{3})^{3/4}\cdot \frac{3}{2} c g^2
(T)T+\frac{T}{\frac{3}{2}c(\frac{a}{3})^{3/4} }
\ee
 where $a\simeq 1.74$ and this justifies our nonperturbative treatment
of the GSM in the temperature region
concerned.

We note that the  transition between these two regimes (with  the
dominance of nonperturbative and then perturbative dynamics) is
smooth,  with the GSM tending to twice the Debye mass at large $T$ [18].

 6. In conclusion, we have pursued a selfconsistent analysis of
 nonperturbative dynamics at $T>T_c$, including the calculation of
 $T_c$, spatial area law and meson and glueball screening masses.
 We stress that in both meson and glueball cases, the nonperturbative
 mechanism plays an essential role in the formation  of the
 screening masses in the temperature region  $0<T<2\div 3 T_c$,
 while at large $T$, the lowest screening masses tend  to the
 corresponding limits, governed by perturbative
 dynamics. We obtain in all cases a reasonable agreement with known
 lattice data supporting our picture of magnetic confinement
 above $T_c$, while there is electric confinement below $T_c$. An
 additional piece of data is necessary to confirm our picture, e.g.
 lattice calculation of the screening glueball wave function to
 compare with our prediction.

One of us (E.G.) is grateful to B.L. Ioffe and V. Eletskii for useful
discussions. The authors are grateful to D. Melikhov for help in
computations. This work was supported by the Russian Fund for Fundamental
Research, grant 93-02-14937.

\begin{figure}[1]
\epsffile{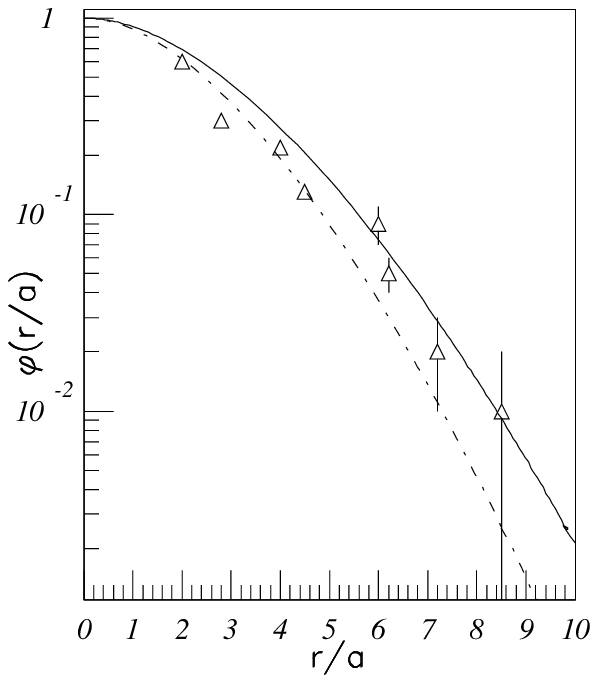}
\caption{Wave
functions for the $\rho$--meson (solid line) and glueball (dashed
line)  vs $r/a$ ($a$=0.23 fm) for $T=210$ MeV. The data
from ref.[5] correspond to $T\approx 210$ MeV.}
\end{figure}

\begin{figure}[2]
\epsffile{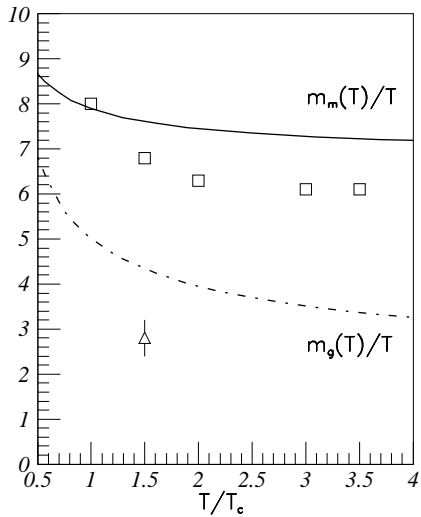}
\caption{The screening masses for mesons (solid line) and glueballs
(dashed line) as a function of temperature. The data from refs.[3,4] (squares)
and [6] (triangle). }
\end{figure}

\end{document}